\documentstyle[preprint,aps]{revtex}
\tighten
\draft
\widetext
\preprint{YITP-00-28}
\bigskip
\bigskip
\begin{document}
\title{Why the Cosmological Constant Problem is Hard}
\medskip
\author{Zurab Kakushadze\footnote{E-mail: 
zurab@insti.physics.sunysb.edu}}
\bigskip
\address{C.N. Yang Institute for Theoretical Physics\\ 
State University of New York, Stony Brook, NY 11794}

\date{June 8, 2000}
\bigskip
\medskip
\maketitle

\begin{abstract} 
{}We consider a recent proposal to solve the cosmological constant problem
within the context of brane world scenarios with infinite volume extra 
dimensions. In such theories bulk can be supersymmetric even if brane
supersymmetry is completely broken. The bulk cosmological constant can 
therefore naturally be zero. Since the volume of the extra dimensions is
infinite, it might appear that at large distances one would measure the
bulk cosmological constant which vanishes. We point out a caveat in this
argument. In particular, we use a concrete model, which is a generalization
of the Dvali-Gabadadze-Porrati model, to argue that in the presence of non-zero
brane cosmological constant at large distances
such a theory might become effectively four dimensional. This is due to a mass
gap in the spectrum of bulk graviton modes. In fact, the corresponding
distance scale is set precisely by the brane cosmological constant. This
phenomenon appears to be responsible for the fact that bulk supersymmetry
does not actually protect the brane cosmological constant.
\end{abstract}

\section{The Model}

{}Recently it was pointed out in
\cite{DGP0,witten} that, in theories where extra dimensions
transverse to a brane have infinite volume\cite{GRS,CEH,DVALI,DGP1,DGP}, 
the cosmological constant on the
brane might be under control even if brane supersymmetry is completely
broken. The key point here is that even if supersymmetry breaking on the
brane does take place, it will not be transmitted to the bulk as the volume 
of the extra dimensions is infinite \cite{DGP0,witten}. Thus, at least in 
principle, we should be able to control some of the properties of the bulk
with the unbroken bulk supersymmetry. In particular, vanishing of the
bulk cosmological constant need not be unnatural\footnote{Note that {\em a
priori} we could have negative cosmological constant consistent with
bulk supersymmetry. However, in the presence of supersymmetry various ways 
are known for ensuring vanishing of the bulk cosmological constant without any
fine-tuning.}.

{}Then the ``zeroth-order'' argument goes as follows \cite{DGP0,witten}.
Let us for definiteness focus on the case of the codimension one brane
embedded in $D$-dimensional space-time. At
least naively, at large (enough) distances, which are precisely relevant 
for the discussion of the cosmological constant, the theory is expected 
to become 
$D$-dimensional. In particular, the laws of gravity, such as Newton's law, 
are expected to become $D$-dimensional at such distances. If so, a brane 
world observer would then really be
measuring the $D$-dimensional (and not $(D-1)$-dimensional) cosmological
constant, which vanishes by bulk supersymmetry. One therefore might 
expect that
the cosmological constant on the brane might somehow also vanish
regardless of brane supersymmetry.

{}The above argument might {\em a priori} have (at least) 
two possible caveats. First, it is
not completely clear what is the relation between the $(D-1)$-dimensional
cosmological constant and the $D$-dimensional one. More precisely, one would
like to see a bit more explicitly how bulk supersymmetry controls the
cosmological constant on the brane. Note that the latter is certainly 
well-defined as being proportional to the curvature on the
brane. In the following we will argue that bulk supersymmetry in a concrete
model of the aforementioned type does not
actually imply vanishing of the brane cosmological constant.
Second, even though the extra dimension has infinite volume, 
{\em a priori} it is not completely obvious why the theory should remain
$D$-dimensional above some large crossover distance scale $r_0$. Thus, one
can imagine a scenario where the theory is effectively $(D-1)$-dimensional
at length scales $r\ll r_0$, it becomes $D$-dimensional at intermediate
scales $r_0\ll r\ll r_*$, and it then again becomes $(D-1)$-dimensional
at larger scales $r_*\ll r$. If so, the natural bound 
(in the General Relativity conventions)  for the cosmological
constant ${\widetilde \Lambda}$ on the brane would be $|{\widetilde \Lambda}|
{\ \lower-1.2pt\vbox{\hbox{\rlap{$<$}\lower5pt\vbox{\hbox{$\sim$}}}}\ } 
1/r_*^2$. As we will argue in the following, this is precisely 
what appears to be the
case in the model we discuss in this paper\footnote{While 
finishing this manuscript, we became aware of the
paper \cite{kogan} where such a phenomenon was argued to occur in a different
brane world model (and in a different context) 
which contains two negative as well as two positive tension
branes.}. In fact, as we will see, in, say, the Dvali-Gabadadze-Porrati model
\cite{DGP}
there is a mass gap in the spectrum of bulk graviton modes
if the cosmological constant on the brane is positive. This then explains
how come the aforementioned ``zeroth-order'' argument does not really apply
as the theory is effectively $(D-1)$-dimensional in the infra-red.

{}In this paper will illustrate the
points raised above in a concrete model. The action for this model is given by:
\begin{equation}\label{action}
 S={\widehat M}_P^{D-3}\int_\Sigma d^{D-1} x \sqrt{-{\widehat G}}\left[
 {\widehat R}-{\widehat\Lambda}\right] +
 M_P^{D-2}
 \int d^D x \sqrt{-G} \left\{R+\zeta\left[R^2-4R_{MN}^2+R_{MNST}^2\right]
 \right\}~.
\end{equation}
For calculational convenience we will keep the number of space-time
dimensions $D$ unspecified.
In (\ref{action}) ${\widehat M}_P$ is (up to a normalization factor - see 
below) the $(D-1)$-dimensional (reduced) Planck scale, while $M_P$ is the 
$D$-dimensional one. The $(D-1)$-dimensional hypersurface $\Sigma$, which we
will refer to as the brane, is the $y=y_0$ slice of the $D$-dimensional 
space-time,
where $y\equiv x^D$, and $y_0$ is a constant. Next, 
\begin{equation}
 {\widehat G}_{\mu\nu}\equiv{\delta_\mu}^M {\delta_\nu}^N G_{MN}
 \Big|_{y=y_0}~,
\end{equation} 
where the capital Latin indices $M,N,\dots=1,\dots,D$, while the Greek
indices $\mu,\nu,\dots=1,\dots,(D-1)$. The quantity ${\widehat\Lambda}$ is
the brane tension. More precisely, there might be various (massless and/or 
massive) fields (such
as scalars, fermions, gauge vector bosons, {\em etc.}), which we
will collectively denote via $\Phi^i$, localized on the brane. Then ${\widehat
\Lambda}={\widehat\Lambda}(\Phi^i,\nabla_\mu\Phi^i,\dots)$ generally depends
on the vacuum expectation values of these fields as well as their derivatives.
In the following we will assume that the expectation values of the $\Phi^i$
fields are dynamically determined, independent of the coordinates 
$x^\mu$, and consistent with $(D-1)$-dimensional general covariance. 
The quantity ${\widehat\Lambda}$ is then a constant which we identify 
as the brane tension. Finally, the coefficient $\zeta$ has the dimension of
length squared, and the term it multiplies is the Gauss-Bonnet term, which is
quadratic in curvature. Also, note that we have set the $D$-dimensional 
bulk cosmological constant to zero, and there are no bulk fields other 
than gravity.

{}The model defined in (\ref{action}) is a generalization of the 
Dvali-Gabadadze-Porrati model recently proposed in \cite{DGP}. In fact, the
difference between the two models (on top of the straightforward 
generalization that we do not 
{\em a priori} assume that the brane is tensionless) 
is the presence of the bulk Gauss-Bonnet term, which we have added in order
to illustrate that higher derivative terms do not seem to modify our
conclusions. The model defined in (\ref{action})
reduces to the Dvali-Gabadadze-Porrati model for $\zeta=0$.

{}Before we turn to our main point, let us briefly comment on the 
$\sqrt{-{\widehat G}} {\widehat R}$ term in the brane world-volume action.
Typically such a term is not included in discussions of various brane world
scenarios (albeit usually the $-\sqrt{-{\widehat G}} {\widehat \Lambda}$ 
term is included). 
However, as was pointed out in \cite{DGP}, even if such a term is absent at the
tree level, as long as the 
brane world-volume theory is not conformal, it will typically be
generated by quantum loops of other fields localized on the 
brane\footnote{This is an important observation which might
sometimes modify various conclusions, and should
in principle be taken into account when discussing other brane world scenarios
as well, for instance, in the Randall-Sundrum type of models \cite{RS}.} 
(albeit 
not necessarily with the desired sign, which, nonetheless, appears to be as
generic as the opposite one).  

{}An important feature of the above model is that (for the standard values of
$D$) we can supersymmetrize the
bulk action. Thus, in the following we will assume that the bulk is
supersymmetric, and the bulk cosmological
constant vanishes. On the other hand, as we have already mentioned, since 
the volume of the $y$ dimension is infinite, supersymmetry on the brane could 
be completely broken, while bulk supersymmetry is intact. In the remainder
of this paper we will address the question whether bulk supersymmetry protects
the brane cosmological constant.

\section{Bulk Supersymmetry and Brane Cosmological Constant}

{}To proceed further, we will need equations of motion following from the
action (\ref{action}). Here we are interested in studying possible solutions
to these equations which are consistent with $(D-1)$-dimensional general 
covariance. That is, we will be looking for solutions with the warped
metric of the following form:
\begin{equation}\label{warped}
 ds_D^2=\exp(2A)ds_{D-1}^2+dy^2~,
\end{equation}
where the warp factor $A$, which is a function of $y$,
is independent of the coordinates
$x^\mu$, and the $(D-1)$-dimensional interval is given
by
\begin{equation}\label{D-1}
 ds_{D-1}^2={\widetilde g}_{\mu\nu}dx^\mu dx^\nu~,
\end{equation}  
with the $(D-1)$-dimensional metric 
${\widetilde g}_{\mu\nu}$ independent of $y$. With this ans{\"a}tz the
equations of motion are given by:
\begin{eqnarray}
 &&\left\{(D-1)(D-2)(A^\prime)^2-
 {{D-1}\over{D-3}}{\widetilde \Lambda}\exp(-2A)\right\}-
 (D-3)(D-4)\zeta\times\nonumber\\
 &&~~~~~~~ \times\Big\{(D-1)(D-2)(A^\prime)^4-2
 {{D-1}\over{D-3}}{\widetilde \Lambda}(A^\prime)^2 \exp(-2A)+\nonumber\\
 \label{A'}
 &&~~~~~~~+
 {{\widetilde\chi}\over{(D-3)(D-4)}}\exp(-4A)\Big\}=0~,\\
 &&\left\{(D-1)(D-2)A^{\prime\prime}+{{D-1}\over {D-3}}
 {\widetilde \Lambda}\exp(-2A)\right\}-2(D-3)(D-4)\zeta\times\nonumber\\
 &&~~~~~~~\times \Big\{
 (D-1)(D-2)A^{\prime\prime}(A^\prime)^2-
 {{D-1}\over{D-3}}{\widetilde \Lambda}\left[A^{\prime\prime}-(A^\prime)^2
 \right]\exp(-2A)-\nonumber\\
 &&~~~~~~~-{{\widetilde\chi}\over{(D-3)(D-4)}}\exp(-4A)
 \Big\}+\nonumber\\
 \label{A''}
 &&~~~~~~~+{{D-1}\over 2}L\left[{\widehat\Lambda}-{\widetilde\Lambda}\exp(-2A)
 \right]\delta(y-y_0)=0~,
\end{eqnarray}
where
\begin{equation}
 L\equiv {\widehat M}_P^{D-3}/M_P^{D-2}~.
\end{equation}
Here ${\widetilde \Lambda}$ is independent of $x^\mu$ and $y$. In fact, it 
is nothing but the cosmological constant of the $(D-1)$-dimensional manifold,
which is therefore an Einstein manifold, corresponding to the hypersurface
$\Sigma$. Our normalization of ${\widetilde\Lambda}$ is such that
the $(D-1)$-dimensional metric ${\widetilde g}_{\mu\nu}$ satisfies
Einstein's equations:
\begin{equation}
 {\widetilde R}_{\mu\nu}-{1\over 2}{\widetilde g}_{\mu\nu}
 {\widetilde R}=-{1\over 2}
{\widetilde g}_{\mu\nu}{\widetilde\Lambda}~,
\end{equation}
so that the $(D-1)$-dimensional Ricci scalar is given by
\begin{equation}
 {\widetilde R}={{D-1}\over{D-3}}{\widetilde\Lambda}~.
\end{equation}
Moreover, the quantity
\begin{equation}
 {\widetilde\chi}\equiv{\widetilde R}^2-4{\widetilde R}_{\mu\nu}^2+
 {\widetilde R}_{\mu\nu\sigma\tau}^2
\end{equation}
is also a constant (if $\zeta\not=0$).

{}Here we note that in the bulk (that is, for $y\not=y_0$) the second
order equation (\ref{A''}) is automatically satisfied once the first
order equation (\ref{A'}) is satisfied. As usual, this is a consequence of
Bianchi identities. 

{}For our purposes here it will not be necessary to find the most general
solutions to the above equations. It will instead suffice to understand
what are the restrictions on the warp factor coming from the requirement
that the bulk be supersymmetric.

\subsection{Killing Spinors and Bulk Supersymmetry}

{}For the bulk to be supersymmetric, we must have covariantly constant
Killing spinors satisfying the following equation (which comes from the
requirement that the bulk gravitino $\psi_M$ have a vanishing variation
under the corresponding supersymmetry transformation):
\begin{equation}
 {\cal D}_M \varepsilon=0~.
\end{equation}
Here $\varepsilon$ is the Killing spinor, and ${\cal D}_M$ is the covariant
derivative
\begin{equation}
 {\cal D}_M\equiv \partial_M +{1\over 4}\Gamma_{AB}{\omega^{AB}}_M~.
\end{equation}
The spin connection ${\omega^{AB}}_M$ is defined via the vielbeins
${e^A}_M$ in the usual way (here the capital Latin indices $A,B,\dots=1,\dots,
D$ are lowered and raised with the $D$-dimensional Minkowski metric
$\eta_{AB}$ and its inverse, while
the capital Latin indices $M,N,\dots=1,\dots,D$ are lowered and raised with
the $D$-dimensional metric $G_{MN}$ and its inverse). Furthermore,
\begin{equation}
 \Gamma_{AB}\equiv{1\over 2}\left[\Gamma_A~,~\Gamma_B\right]~,
\end{equation} 
where $\Gamma_A$ are the constant Dirac gamma matrices satisfying
\begin{equation}
 \left\{\Gamma_A~,~\Gamma_B\right\}=2 \eta_{AB}~.
\end{equation}

{}Next, we would like to study the above Killing spinor equations in the
warped backgrounds of the form (\ref{warped}):
\begin{eqnarray}
 \label{Kill2}
 &&\varepsilon^\prime=0~,\\
 \label{Kill3}
 &&{\widetilde {\cal D}}_\mu \varepsilon+{1\over 2}A^\prime \exp(A)
 {\widetilde \Gamma}_\mu \Gamma_D\varepsilon=0~.
\end{eqnarray}
Here ${\widetilde {\cal D}}_\mu$ is the $(D-1)$-dimensional covariant
derivative corresponding to the metric ${\widetilde g}_{\mu\nu}$,
${\widetilde \Gamma}_\mu$ are the $(D-1)$-dimensional Dirac gamma
matrices satisfying
\begin{equation}
 \left\{{\widetilde \Gamma}_\mu~,~{\widetilde \Gamma}_\nu\right\}=
 2{\widetilde g}_{\mu\nu}~.
\end{equation}
Also, note that $\Gamma_D$, which is the $D$-dimensional Dirac
gamma matrix $\Gamma_M$ with $M=D$ (that is, the Dirac gamma matrix
corresponding to the $x^D=y$ direction) is constant in this background.

{}To begin with, note that (\ref{Kill2}) and (\ref{Kill3}) do not have
a solution unless 
\begin{equation}
 A^\prime\exp(A)=C~,
\end{equation}
where $C$ is some constant. Let us assume that this condition is satisfied.
We can rewrite the system of equations (\ref{Kill2}) and (\ref{Kill3}) as 
follows. Let
\begin{equation}
 \gamma_\mu\equiv{\widetilde \Gamma}_\mu\Gamma_D~.
\end{equation}
These new gamma matrices satisfy
\begin{equation}
 \left\{\gamma_\mu~,~\gamma_\nu\right\}=-2{\widetilde g}_{\mu\nu}\equiv 2
 \rho_{\mu\nu}~,
\end{equation}
that is, $\gamma_\mu$ are the gamma matrices for a space with the metric
\begin{equation}
 \rho_{\mu\nu}=-{\widetilde g}_{\mu\nu}~,
\end{equation}
whose signature is $(+,-,\dots,-)$. The Killing spinor equations now read
(note that the covariant derivative ${\widetilde {\cal D}}_\mu$ is unaffected
by the metric inversion)
\begin{eqnarray}
 \label{Kill2'}
 &&\varepsilon^\prime=0~,\\
 \label{Kill3'}
 &&{\widetilde {\cal D}}_\mu \varepsilon+{1\over 2}C
 \gamma_\mu \varepsilon=0~,
\end{eqnarray}
which have non-trivial solutions for 
the ${\rm AdS}_{D-1}\times {\bf R}$ space with the
signature $(+,-,\dots,-)$ and negative cosmological constant for the 
${\rm AdS}_{D-1}$ piece given by
\begin{equation}
 \lambda=-(D-2)(D-3)C^2~.
\end{equation}
Note that under the metric inversion the Ricci scalar and, therefore,
the cosmological constant flip their sign. This implies that the Killing spinor
equations have a non-trivial solution provided that the metric ${\widetilde
g}_{\mu\nu}$ on the brane corresponds to a de Sitter space with the signature
$(-,+,\dots,+)$ and positive cosmological constant
\begin{equation}\label{susy}
 {\widetilde\Lambda}=-\lambda=(D-2)(D-3)C^2=(D-2)(D-3)(A^\prime)^2\exp(2A)~.
\end{equation}
Here we note that for such warp factors the bulk curvature, which is
given by
\begin{equation}
 R={\widetilde R}\exp(-2A)-(D-1)\left[2A^{\prime\prime}+D(A^\prime)^2\right]~,
\end{equation}
vanishes.

{}Next, the fact that the metric $\rho_{\mu\nu}$ is that of the 
${\rm AdS}_{D-1}$ space (with the signature $(+,-,\dots,-)$) implies that
the corresponding Riemann tensor
\begin{equation}
 {\cal R}_{\mu\nu\sigma\tau}={\lambda\over{(D-2)(D-3)}}\left[
 \rho_{\mu\sigma}\rho_{\nu\tau}-\rho_{\mu\tau}
 \rho_{\nu\sigma}\right]~.
\end{equation}
Since the Riemann tensor flips sign under the metric inversion, we obtain
\begin{equation}
 {\widetilde R}_{\mu\nu\sigma\tau}=
 {{\widetilde \Lambda}\over{(D-2)(D-3)}}\left[
 {\widetilde g}_{\mu\sigma}{\widetilde g}_{\nu\tau}-{\widetilde g}_{\mu\tau}
 {\widetilde g}_{\nu\sigma}\right]~.
\end{equation} 
This implies that ${\widetilde \chi}$ defined above is given by
\begin{equation}\label{chi}
 {\widetilde \chi}={{(D-1)(D-4)}\over{(D-2)(D-3)}}{\widetilde\Lambda}^2~.
\end{equation}

{}Now we can readily see that (\ref{A'}), as well as (\ref{A''}) in the
bulk, are satisfied as long as we have bulk supersymmetry, which implies 
(\ref{susy}) and ({\ref{chi}). As to the brane cosmological constant
${\widetilde\Lambda}$, it is related to the brane tension ${\widehat\Lambda}$
via the jump condition which follows from (\ref{A''}). The important point here
is that it does not have to be zero to preserve bulk supersymmetry. 

\subsection{The Dvali-Gabadadze-Porrati Model}

{}The last result seems to indicate that the aforementioned ``zeroth-order''
argument must somehow break down in the above model. Here we would like to
better understand the precise mechanism for this breakdown. For simplicity
we will do this in the Dvali-Gabadadze-Porrati model, that is, for
$\zeta=0$. Then for positive ${\widetilde \Lambda}$
we have the following bulk equation
\begin{equation}
 (A^\prime)^2\exp(2A)=C^2~.
\end{equation}
Taking into account the jump condition which follows from (\ref{A''}), we
obtain the following non-singular solution with infinite volume:
\begin{eqnarray}
 &&A(y)=\ln\left[C(y_+-y)\right]~,~~~y<y_0~,\\
 &&A(y)=\ln\left[C(y-y_-)\right]~,~~~y\geq y_0~, 
\end{eqnarray}
where 
\begin{equation}
 y_\pm\equiv y_0\pm\Delta~,
\end{equation} 
and without loss of generality we have
assumed $C>0$. Note that the positive quantity $\Delta$ is fixed from the jump
condition
\begin{equation}
 \left\{{\widehat\Lambda}-{\widetilde\Lambda}\exp\left[-2A(y_0)\right]
 \right\}+2(D-2){1\over L}\left[
 A^\prime(y_0+)-A^\prime(y_0-)\right]=0~,
\end{equation}
which can be rewritten as 
\begin{equation}
 {\widehat\Lambda}=(D-2)(D-3){1\over\Delta^2}-4(D-2){1\over{L\Delta}}~.
\end{equation}
Let us discuss the
possible solutions of this equation for $\Delta$.

{}Thus, if ${\widehat\Lambda}>0$, then the solution is given by (here we are
assuming $L>0$):
\begin{equation}
 \Delta={{2(D-2)}\over{{\widehat\Lambda}L}}\left[\sqrt{1+{{D-3}\over{D-2}}
 {\widehat\Lambda}L^2}-1\right]~.
\end{equation} 
On the other hand, if ${\widehat\Lambda}<0$, then we must have
\begin{equation}
 |{\widehat\Lambda}|\leq{{D-2}\over{D-3}}{1\over L^2}~.
\end{equation}
We then have two solutions
\begin{equation}
 \Delta_\pm={{2(D-2)}\over{|{\widehat\Lambda}|L}}\left[1\pm 
 \sqrt{1-{{D-3}\over{D-2}}
 |{\widehat\Lambda}| L^2}~\right]~.
\end{equation}
Thus, we have a lower bound on the brane tension.

{}Here the following remark is in order. In the above solution the
effective brane tension, defined as
\begin{equation}
 f\equiv {\widehat\Lambda}-{\widetilde\Lambda}\exp[-2A(y_0)]~,
\end{equation}
is negative. Such a brane would suffer from world-volume ghosts unless we
assume that it is an ``end-of-the-world'' brane located at an orbifold
fixed point. Thus, in the above solution the geometry of the $y$ dimension
is that of ${\bf R}/{\bf Z}_2$ (and not of ${\bf R}$), with the orbifold
fixed point identified with $y_0$ (note that the above solution has the
${\bf Z}_2$ symmetry required for the orbifold interpretation), and the
brane is stuck at the orbifold fixed point. Note that there is another
solution with positive effective brane tension\footnote{Since $f>0$ in this
case, {\em a priori} there is no need to restrict to the orbifold
interpretation even though the solution does possess the corresponding
${\bf Z}_2$ symmetry.}, which is given by
\begin{eqnarray}
 &&A(y)=\ln\left[C(y-y_-)\right]~,~~~y_-<y<y_0~,\\
 &&A(y)=\ln\left[C(y_+ -y)\right]~,~~~y_+>y\geq y_0~. 
\end{eqnarray}
In this solution\footnote{A $D=4$ version of this solution can
be found in \cite{Vilenkin}.} the volume of the extra dimension is finite as
the latter is cut off by horizons located at $y=y_\pm$. Note that in
such a solution the argument of \cite{DGP0,witten} does not apply to begin
with, and this is precisely why we focus on the above solution
with negative effective brane
tension\footnote{Actually, there is the third solution with
vanishing effective brane
tension $f=0$, which is given (up to equivalence under
the reflection around $y_0$) by $A(y)=\ln[C(y-y_-)]$, $y>y_-$. In this
solution the space in the $y$ direction is cut off by a horizon at $y=y_-$.
As far as our discussion in this paper is concerned, this solution has 
essentially the same properties as the one with negative effective brane
tension, which we will focus on in the following.}.   

{}Next, note that by rescaling the coordinates $x^\mu$ on the brane we can
always set $\exp[A(y_0)]=1$. This is equivalent to setting $C=1/\Delta$.
In this case the $(D-1)$-dimensional Planck 
scale (which is related to the $(D-1)$-dimensional Newton's constant
arising in the $(D-1)$-dimensional Newton's law) is simply ${\widehat M}_P$, 
while the $D$-dimensional Planck scale is $M_P$. The effective 
$(D-1)$-dimensional cosmological constant (in the Particle Physics conventions)
is given by
\begin{equation}
 {\widetilde\Lambda}_{\rm eff}\equiv{\widetilde\Lambda}{\widehat M}_P^{D-3}~.
\end{equation}
Its ratio to ${\widehat M}_P^{D-1}$ is then ${\widetilde \Lambda}/{\widehat 
M}_P^2\sim 1/\Delta^2 {\widehat M}_P^2$. Let us assume that 
${\widehat \Lambda}>0$ (which is generically expected to be the case
once brane supersymmetry is broken). Then it is not difficult to see that
we have
\begin{equation}
 \Delta{\ \lower-1.2pt\vbox{\hbox{\rlap{$<$}\lower5pt\vbox{\hbox{$\sim$}}}}\ }
 1/\sqrt{\widehat\Lambda}~.
\end{equation}
This then implies that 
\begin{equation}
 {\widetilde\Lambda}/{\widehat M}_P^2
 {\ \lower-1.2pt\vbox{\hbox{\rlap{$>$}\lower5pt\vbox{\hbox{$\sim$}}}}\ }
 {\widehat \Lambda}/{\widehat M}_P^2~.
\end{equation}
That is, the brane cosmological constant is at 
least\footnote{Phenomenologically the crossover scale $L$ is supposed to be
quite large \cite{DGP}, 
so that we actually expect ${\widehat\Lambda}\gg1/L^2$, and
$\Delta\sim1/\sqrt{\widehat\Lambda}$. Because of this, the aforementioned
lower bound on the brane tension in the case of negative ${\widehat\Lambda}$
would seem to require quite a bit of fine-tuning.} 
as large as the brane
tension, which in a four dimensional theory is generically expected
to be $\sim({\rm TeV})^4$ (in the Particle Physics conventions) assuming that
the brane supersymmetry breaking scale is $\sim{\rm TeV}$. 

{}Thus, we have arrived at the same unpleasant generic lower 
bound on the cosmological constant as in the usual four dimensional effective
field theory. Naturally, one would like to better understand how come the
aforementioned ``zeroth-order'' argument does not hold in the 
Dvali-Gabadadze-Porrati model. To do this, we will need to study the bulk 
graviton spectrum in this model.

\begin{center}
 {\em Normalizable Modes}
\end{center}

{}From (\ref{action}) it is not difficult to see that
the norm of a bulk graviton mode is proportional to
\begin{equation}
 ||{\widetilde h}_{\mu\nu}||^2\sim\int dy\exp[(D-3)A]\sigma^2~,
\end{equation}
where $\sigma=\sigma(y)$ depends only on $y$. Moreover, 
$\sigma(y)$ satisfies the following equation \cite{freedman,zk}:
\begin{equation}\label{sigma}
 \left(\exp[(D-1)A]\sigma^\prime\right)^\prime+m^2\exp[(D-3)A]\sigma=0~,
\end{equation}
where $m^2$ is the mass squared of the corresponding graviton mode.

{}To study normalizability of these graviton modes, let us make the coordinate
transformation $y\rightarrow z$ so that the metric takes the form:
\begin{equation}
 ds^2_D=\exp(2A)\left({\widetilde g}_{\mu\nu}dx^\mu dx^\nu+dz^2\right)~.
\end{equation}
That is, 
\begin{equation}
 dy=\exp(A)dz~, 
\end{equation}
where we have chosen the overall sign so that $z\rightarrow\pm\infty$ as
$y\rightarrow\pm\infty$. We will conveniently choose the origin for the
$z$ coordinate to correspond to $y=y_0$. Then we have
\begin{eqnarray}
 &&z=-{1\over C}\ln\left[{{y_+-y}\over\Delta}\right]~,~~~y<y_0\\
 &&z=+{1\over C}\ln\left[{{y-y_-}\over\Delta}\right]~,~~~y\geq y_0~.
\end{eqnarray} 
We then have
\begin{equation}\label{A(z)}
 A(z)=C|z|+\ln(C\Delta)~.
\end{equation}
In terms of the $z$ coordinate we have
\begin{equation}
 \sigma_{zz}+(D-2)A_z\sigma_z+m^2\sigma=0~,
\end{equation}
where the subscript $z$ denotes derivative w.r.t. $z$. Let
\begin{equation}
 \sigma\equiv \exp\left[-{1\over 2}(D-2) A\right]{\widehat \sigma}~.
\end{equation}
Then the norm of ${\widetilde h}_{\mu\nu}$ is given by
\begin{equation}
 ||{\widetilde h}_{\mu\nu}||^2\sim\int dz {\widehat \sigma}^2~.
\end{equation}
This implies that for a given mode to be plane-wave/quadratically 
normalizable, ${\widehat\sigma}$ must be plane-wave/quadratically
normalizable w.r.t. the flat $z$ coordinate.

{}The equation for ${\widehat\sigma}$ reads:
\begin{equation}\label{sigmahat}
 {\widehat\sigma}_{zz}+\left[m^2 -{1\over 2}(D-2) A_{zz}-{1\over 4}
 (D-2)^2 (A_z)^2\right]{\widehat\sigma}=0~.
\end{equation}
Using (\ref{A(z)}) we obtain
\begin{equation}\label{sigmahat1}
 {\widehat\sigma}_{zz}+\left[\mu^2 -2m_*\delta(z)\right]
 {\widehat\sigma}=0~,
\end{equation}
where 
\begin{equation}
 \mu^2\equiv m^2-m^2_*~,
\end{equation}
and
\begin{equation}
 m_*\equiv {1\over 2}(D-2)C~.
\end{equation}
For $m^2>m_*^2$ the solution of (\ref{sigmahat1}) reads
\begin{equation}
 {\widehat \sigma}(z)={\rm const.}\times\left[\cos(\mu z)+{m_*\over\mu}
 \sin(\mu|z|)\right]~.
\end{equation}
Thus, the bulk gravitons with $m^2>m_*^2$ are plane-wave normalizable. For
$m^2=m_*^2$ the solution reads:
\begin{equation}
 {\widehat\sigma}(z)={\rm const.}\times \left[1+m_*|z|\right]~,
\end{equation}
so that this mode is not normalizable. Finally, for $m^2<m_*^2$ the solution
is given by
\begin{equation}
 {\widehat\sigma}(z)={\rm const.}\times\left[\cosh\left(\sqrt{-\mu^2} z\right)
 +{m_*\over\sqrt{-\mu^2}}\sinh\left(\sqrt{-\mu^2}|z|\right)\right]~, 
\end{equation}
which is not plane-wave normalizable. Thus, we see that if the cosmological
constant on the brane is positive, then we have a {\em mass gap} in the 
spectrum of the bulk modes. Once again, let us set $C=1/\Delta$ (so that the
$(D-1)$-dimensional Planck scale is given by ${\widehat M}_P$). Then we have
plane-wave normalizable bulk gravitons with masses larger than
\begin{equation}
 m_*={{D-2}\over 2}{1\over \Delta}
 {\ \lower-1.2pt\vbox{\hbox{\rlap{$>$}\lower5pt\vbox{\hbox{$\sim$}}}}\ }
 \sqrt{\widehat\Lambda}~.
\end{equation}
That is, the theory is actually $(D-1)$ dimensional at distance scales
$r{\ \lower-1.2pt\vbox{\hbox{\rlap{$>$}\lower5pt\vbox{\hbox{$\sim$}}}}\ }r_*$,
where
\begin{equation}
 r_*\sim 1/m_*~.
\end{equation}

{}Now suppose that $L\ll r_*$. Then at distance scales $
r{\ \lower-1.2pt\vbox{\hbox{\rlap{$<$}\lower5pt\vbox{\hbox{$\sim$}}}}\ }L$ the
theory is effectively $(D-1)$ dimensional, at scales $L
{\ \lower-1.2pt\vbox{\hbox{\rlap{$<$}\lower5pt\vbox{\hbox{$\sim$}}}}\ }
r {\ \lower-1.2pt\vbox{\hbox{\rlap{$<$}\lower5pt\vbox{\hbox{$\sim$}}}}\ }r_*$
the theory is $D$-dimensional, and, finally, at scales $r
{\ \lower-1.2pt\vbox{\hbox{\rlap{$>$}\lower5pt\vbox{\hbox{$\sim$}}}}\ }r_*$
it is $(D-1)$-dimensional again. This explains how come bulk supersymmetry
does not protect the brane cosmological constant - the $(D-1)$-dimensional
effective field theory is a good approximation below the energy scales
$\sim m_*$. On the other hand, if 
$r_*{\ \lower-1.2pt\vbox{\hbox{\rlap{$<$}\lower5pt\vbox{\hbox{$\sim$}}}}\ }L$,
then the theory is never $D$-dimensional but is always
$(D-1)$-dimensional. Note 
that without fine-tuning 
in the phenomenological context we actually expect $r_*\ll L$.

{}Thus, to obtain a small cosmological constant on the brane we must fine-tune
the brane tension. Note that in such a brane world model the present day
cosmological 
evolution on the brane at scales comparable with the size of our Universe
would be described by the four dimensional laws of gravity. In fact, this
is expected to be the case even at earlier evolutionary stages such as 
inflation. On the other hand, the five dimensional nature of the model would
have to show up at somewhat lower scales of order of the crossover distance
scale $L$. 

\acknowledgments

{}I would like to thank Gia Dvali, Gregory Gabadadze,
Martin Ro{\v c}ek, Tom Taylor and Stefan Vandoren
for discussions.
This work was supported in part by the National Science Foundation.
I would also like to thank Albert and Ribena Yu for financial support.


\begin{references}

\bibitem{DGP0} G. Dvali, G. Gabadadze and M. Porrati, hep-th/0002190.

\bibitem{witten} E. Witten, hep-ph/0002297.

\bibitem{GRS} R. Gregory, V.A. Rubakov, S.M. Sibiryakov, hep-th/0002072;
hep-th/0003045.

\bibitem{CEH} C. Csaki, J. Erlich and T.J. Hollowood, hep-th/0002161;
hep-th/0003020.

\bibitem{DVALI} G. Dvali, hep-th/0004057.

\bibitem{DGP1} G. Dvali, G. Gabadadze and M. Porrati, hep-th/0003054.

\bibitem{DGP} G. Dvali, G. Gabadadze and M. Porrati, hep-th/0005016.

\bibitem{RS} L. Randall and R. Sundrum, Phys. Rev. Lett. {\bf 83} (1999) 3370;
Phys. Rev. Lett. {\bf 83} (1999) 4690.

\bibitem{kogan} I.I. Kogan, S. Mouslopoulos, A. Papazoglou, G.G. Ross, 
hep-th/0006030.

\bibitem{Vilenkin} J. Ipser and P. Sikivie, Phys. Rev. {\bf D30} (1984) 712;\\
A. Vilenkin, Phys. Lett. {\bf B133} (1983) 177.

\bibitem{freedman} O. DeWolfe, D.Z. Freedman, S.S. Gubser and A. Karch,
hep-th/9909134.

\bibitem{zk} Z. Kakushadze, hep-th/0005217.



\end{references}
\end{document}